\documentclass[graybox]{svmult}

\usepackage{mathptmx}       % selects Times Roman as basic font
\usepackage{helvet}         % selects Helvetica as sans-serif font
\usepackage{courier}        % selects Courier as typewriter font
\usepackage{type1cm}        % activate if the above 3 fonts are
\usepackage{makeidx}         % allows index generation
\usepackage{graphicx}        % standard LaTeX graphics tool
\usepackage{multicol}        % used for the two-column index
\usepackage[bottom]{footmisc}% places footnotes at page bottom

\makeindex             % used for the subject index
\begin{document}

\title*{Black-Hole Lattices}
% Use \titlerunning{Short Title} for an abbreviated version of
% your contribution title if the original one is too long
\author{Eloisa Bentivegna}
% Use \authorrunning{Short Title} for an abbreviated version of
% your contribution title if the original one is too long
\institute{Eloisa Bentivegna \at Max-Planck-Institut f\"ur Gravitationsphysik 
(Albert-Einstein-Institut), Am M\"uhlenberg 1, D-14476 Golm, Germany,
\email{eloisa.bentivegna@aei.mpg.de}}
%
% Use the package "url.sty" to avoid
% problems with special characters
% used in your e-mail or web address
%
\maketitle

\abstract{The construction of black-hole lattices, first 
attempted by Richard Lindquist and John Wheeler in 1957, has
recently been tackled with renewed interest, as a test bed for studying
the behavior of inhomogeneities
in the context of the backreaction problem.
In this contribution, I discuss how black-hole lattices can help 
shed light on two important issues, and illustrate the conclusions reached
so far in the study of these systems.}

\section{Introduction}
The first appearance of the concept of a periodic arrangement of 
black holes can be found in~\cite{RevModPhys.29.432}. There, the authors discuss
a strategy to stitch together patches of the Schwarzschild solution
so as to construct a space with a discrete translational symmetry but some
degree of spatial inhomogeneity.

In their work, the stitching prescription does not
lead to a global solution of Einstein's equation. Accepting the
constraint violations, however, buys one some freedom in the specification
of such prescription, which the authors use to impose that the
time evolution of a suitably-defined scale factor in this space
follows that of a universe filled with dust of the same total mass.
One then has a simple, analytical test bed in which to measure the effect
of inhomogeneities in, say, the optical properties of a cosmological
model. In this work and in subsequent ones~\cite{springerlink:10.1007/BF01889418}, it was pointed
out how an exact initial-data construction could be obtained.

A few years ago, Clifton and Ferreira extended this model, originally
limited to the positive-curvature case, to zero and negative curvature~\cite{Clifton:2009jw}. Again, the
junction conditions were designed to reproduce an assigned time evolution, 
and the models were used to explore the propagation of null rays in
an inhomogeneous universe.

Recently, the first exact initial data describing a black-hole lattice
have been analyzed~\cite{Clifton:2012qh} and evolved in full numerical 
relativity~\cite{Bentivegna:2012ei}. This has helped make progress on two fronts:
\begin{enumerate}
\item From a conceptual point of view, it has clarified some of the
conditions under which black-hole solutions can be glued together;
this gives some insight into the requirements for constructing a metric tensor for the 
universe starting from the basic building block of a spherically-symmetric,
isolated object.
It turns out that these conditions are remarkably close to the 
conditions for the existence of homogeneous, periodic solutions of Einstein's
equation. The requirements that periodic boundary conditions impose
on the Hamiltonian constraint are likely at the root of this 
correspondence.
\item From a practical standpoint, the time evolution of a
lattice gives one example of the behavior of inhomogeneities
in a cosmological setting and in the non-linear regime, thereby 
serving as a nice complement to perturbative studies and
the averaging framework.
Surprisingly, even the time development of these lattices
remains in some sense close to the counterpart model in the dust
Friedmann-Lema\^itre-Robertson-Walker (FLRW) class. 
\end{enumerate}
In the following two sections, I will discuss the initial-data
construction and illustrate the time evolution of a 
black-hole lattice with positive conformal curvature.

\section{The Construction of Exact Black-Hole-Lattice Initial Data}
As pointed out in~\cite{RevModPhys.29.432}, in order to construct an exact black-hole 
lattice one should directly tackle the Einstein constraints. Working in the
\emph{conformal transverse-traceless} decomposition, these read:
\begin{eqnarray}
\label{eq:CTTconstraints}
 \tilde \Delta \psi - \frac{\tilde R}{8}\,\psi - \frac{K^2}{12}\,\psi^5 + \frac{1}{8} {\tilde A}_{ij} {\tilde A}^{ij} \psi^{-7} = -2\pi \,\psi^5\,\rho \\
 \tilde D_i \tilde A^{ij} - \frac{2}{3} \psi^6 \tilde \gamma^{ij} \tilde D_i K = 0
\end{eqnarray}
Let us focus on the hamiltonian constraint first. If one integrates this
equation over one of the cells of the black-hole lattice, the following condition is obtained:
\begin{eqnarray}
\label{eq:IntHamiltonianconstaint}
%\nonumber
\int_D &&\left(\frac{\tilde R}{8}\,\psi + \frac{1}{12}\,K^2\,\psi^5 - \frac{1}{8}\psi^{-7}\,\tilde A^{ij}\,\tilde A_{ij} 
\right) \sqrt{\tilde\gamma}\,d^3 x =
2\pi \Sigma_{i=1}^{N} m_i
\end{eqnarray}
where $m_i$ represent the masses of the black holes contained in the cell. This condition implies
that $\tilde R$ and $K_{ij}$ cannot both be zero. In other words, conformally-flat 
lattices do not admit a time-symmetric spatial hypersurface; vice versa, lattices
with a $K=0$ spatial hypersurface must be conformally curved. This mirrors the
identical property of the FLRW class. 

The two simplest roads to the construction of a periodic black-hole lattice are
thus the following:
\begin{itemize}
\item Choosing $K=0$, and solving:
\begin{equation}
 \tilde \Delta \psi - \frac{\tilde R}{8}\,\psi = 0
\end{equation}
Equation (\ref{eq:IntHamiltonianconstaint}) implies that $R>0$, so that the spacetime
can be foliated by conformally-$S^3$ hypersurfaces.
As shown in~\cite{springerlink:10.1007/BF01889418}, this equation can be solved exactly. 
Furthermore, it is linear, so one can simply
superimpose known solutions to generate new ones. Notice, however, that if one is interested 
in regular lattices, only six possible arrangements of black holes are possible,
corresponding to the six regular tessellations of $S^3$, which consist of 
$N=5,8,16,24,120$ and $600$ cells.
\item Choosing $\tilde R = 0$, and solving:
\begin{eqnarray}
 \tilde \Delta \psi - \frac{K^2}{12}\,\psi^5 + \frac{1}{8} {\tilde A}_{ij} {\tilde A}^{ij} \psi^{-7} = 0 \\
 \tilde D_i \tilde A^{ij} - \frac{2}{3} \psi^6 \tilde \gamma^{ij} \tilde D_i K = 0
\end{eqnarray}
This system is more difficult to solve, as the Hamiltonian constraint is non-linear
and the momentum constraint is not an identity as in the previous case. For a numerical
approach to the problem, see~\cite{Yoo:2012jz}.
\end{itemize}

\begin{figure}[t]
\begin{center}
\includegraphics[scale=.2, angle=270]{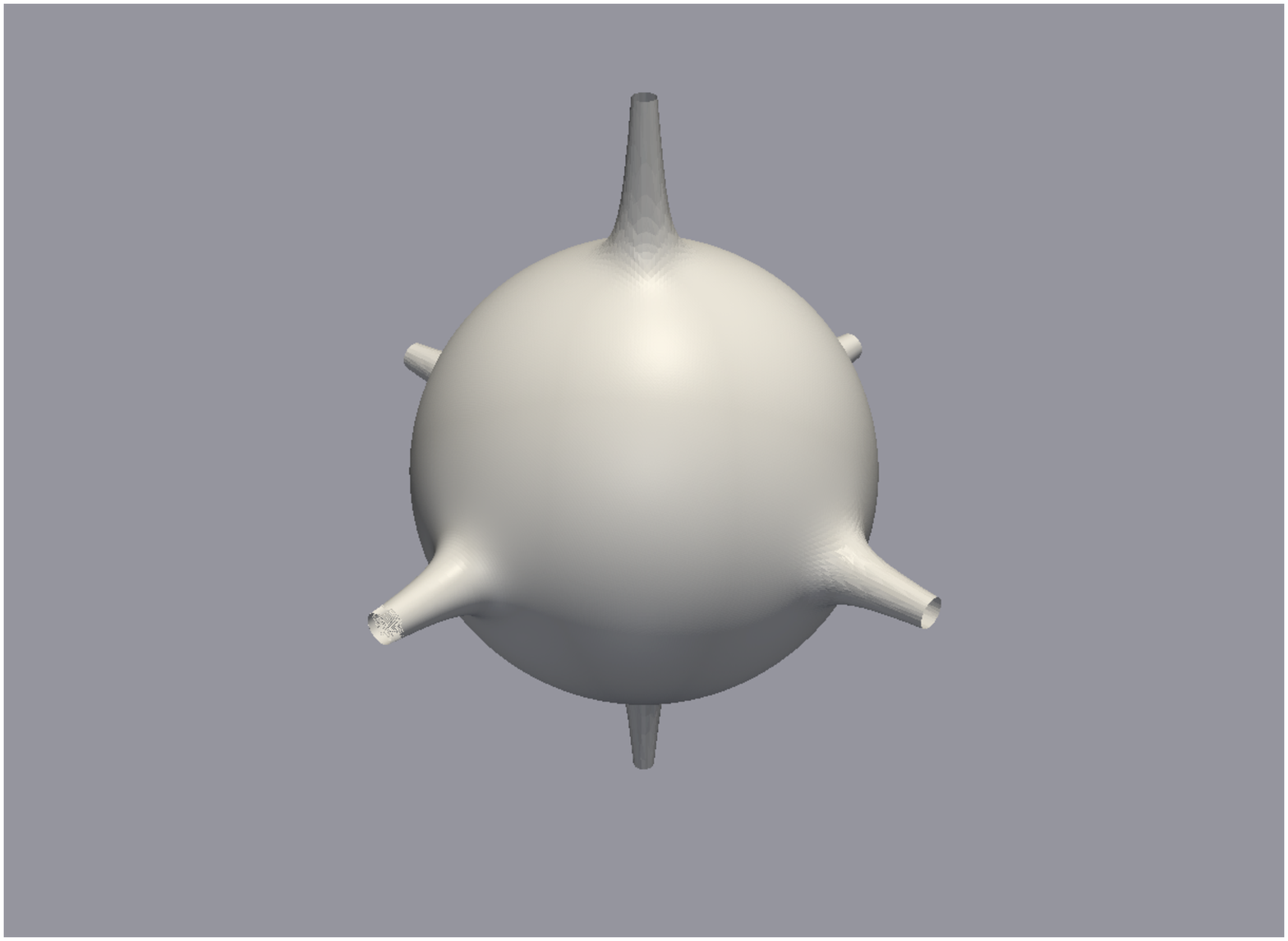}
\caption{A two-dimensional section of the $N=8$ $S^3$ black-hole
lattice, embedded in three dimensions.}
\label{fig:1}
\end{center}
\end{figure}

\section{The Evolution of an $S^3$ Lattice of Eight Black Holes}
\begin{figure}
\begin{center}
\includegraphics[scale=0.9, trim=150 450 150 130, clip=true]{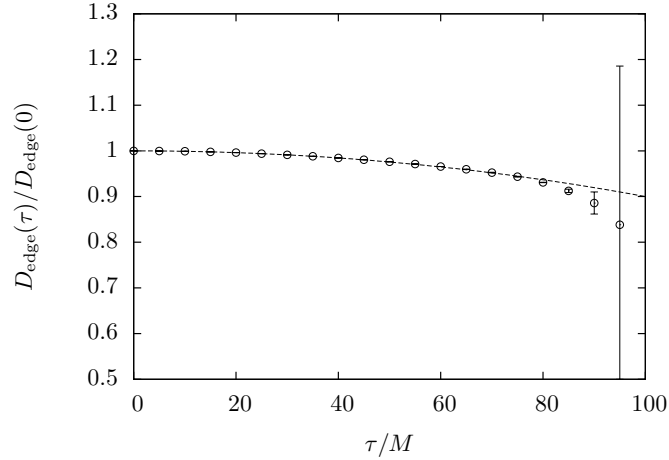}
\caption{Proper length of the edge of a lattice cell as a function
of proper time $\tau$. The dashed line represents a closed FLRW model
in which the edge of a cell of the $N=8$ tessellation is equal to
$D_{\rm edge}(0)$ initially.}
\label{fig:2}
\end{center}
\end{figure}

In~\cite{Bentivegna:2012ei}, the initial data for the $N=8$ $S^3$
lattice (a section of which is shown in Figure \ref{fig:1})
has been evolved in time for approximately one third of the
corresponding FLRW recollapse time. A scale factor can be defined via the proper
length of one of the cell edges; its evolution is shown in Figure \ref{fig:2}.
This scale factor is compatible with the FLRW result in this entire 
time window; eventually, though, due to the gauge condition used to evolve
this system, reaching later and later
values of the proper time is subject to an increasing numerical error,
and eventually becomes impossible.

\begin{acknowledgement}
I acknowledge the financial support from a Marie Curie International Reintegration 
Grant (PIRG05-GA-2009-249290).
\end{acknowledgement}

%\section*{Appendix}
%\addcontentsline{toc}{section}{Appendix}

%\bibliographystyle{plain}
%\bibliography{references}

\end{document}